# Magnetic poles enabled kirigami meta-structure for stable mechanical memory storage with high information density

Libiao Xin[1,4,5], Yanbin Li[3,4,5], Baolong Wang[1,4], Zhiqiang Li[2,5]

1 College of Mechanical and Vehicle Engineering, Taiyuan University of Technology, Taiyuan, 030024, China

2 College of Aeronautics and Astronautics, Taiyuan University of Technology, Taiyuan, 030024, China

3 Department of Mechanical Engineering, North Carolina State University, Raleigh, NC, 27606, USA

4 These authors contribute equally to this work.

5 Corresponding author. Email: xinlibiao@tyut.edu.cn, yli255@ncsu.edu, lizhiqiang@tyut.edu.cn

**Abstract**: Some bi/multi-stable Mechanical meta-structures have been implemented as mechanical memory devices which however are with limits such as complex structural forms, low information storage capability and/or fragile structural stability to maintain the stored information bits robustly under external interferences. To address these issues, we refer to the structural intelligence by constructing a simple 3D-printable multi-layered cylindrical kirigami module with gradient structural parameters and propose a mechanical memory device that can robustly store information bits exponentially larger than previous designs. We demonstrate the promising enhancement of information storage capability of our proposed mechanical memory device relies on two mechanisms: (1) the deformation sequences of the kirigami module enabled by the gradient structural parameter, which brings the extra dimensional degree of freedom to break the traditional mechanical memory unit with only planar form and merits information bits with spatially combinatorical programmability, and (2) the combinatorics of the deformation independences among the cylindrical kiriagmi unit arrays in the constructed mechanical memory device. Particularly, we achieve both the structural stabilities and the desired structural robustness in the mechanical memory devices by additively introducing magnetic "N-S" poles in units, which can protect the stored information from interferences like mechanical crushing, impacting and/or shaking.

## INTRODUCTION

Mechanical meta-structure (or meta-material) have been demonstrated with various exotic properties beyond natural materials (*1-4*). Pre-assigned with unique structural units, mechanical metamaterials can be integrated with mechanical intelligence (such as snap-through based instability (*5, 6*), bi/multi-stability (*7-9*), topological (re)programmability (*10-12*)) and/or materials intelligence (*3, 13, 14*) (for example combining with thermo-/electro-/magneto-actuated materials including liquid crystal elastomer (*15, 16*), shape memory alloy/polymer (*17-19*), ferromagnetic material (*20-23*), hydrogel (*24, 25*), dielectric materials (*26*) etc.) to achieve tunable and (re)programmable mechanical properties (*27, 28*), morph target shapes (*29-31*), imitate electrical circuits (*32, 33*), perform logic computation (*27*), encrypt/process simple information (*34, 35*), sense/intercept external environment conditions (*36*), and/or act as multi-functional robotic structural platforms (*37*). Specifically, given the equivalence between the "0" and "1" memory bits compare with the first and second stable deformation states in bi-stable structures, mechanical meta-structures have been recently to be constructed as mechanical memory storage devices (*38, 39*).

Compared with conventional electrical memory devices (*40, 41*), mechanical meta-structure based memory devices (*38, 39*) can robustly avoid deterioration from external physical/chemical interference, and more importantly prevent malfunction from any destructive electromagnetic invasion. However, existing mechanical meta-structures (*34, 35, 38, 39*) for memory uses are either with limited memory storage capabilities (less memory bits) or easily to be interrupted and unable to maintain the stored information mechanically stable enough (i.e., with small resistance force to battle with external interferences like severe vibration, shaking, stretching or compressing). Intrinsically, there are two ways to improve memory storage capability: (1) enlarging unit numbers in plane, and (2) exploiting the extra spatial dimensional design degree of freedom to construct independently deformable multi-layered units by which the combinatorics of deformed units can be implemented to increase memory bits volume. Moreover, additively beneficial factors from mechanical/material intelligence (*42, 43*) can be considered for memory storage meta-structures to more robustly enhance the structural stability. To the best of the authors knowledge, rare efforts has been conducted to effectively solve the abovementioned two issues simultaneously in one design.

In this work, inspired by the simple beam buckling theory we proposed a magnetic kirigami-based cylindrical meta-structure that can outperform any previous mechanical meta-structures for mechanical memory uses particularly in terms of the storage capability and more importantly the robust and stable maintenance of the stored memory bits. Fig. 1A shows the overall design concept of our work, i.e. a meta-structure based memory box containing $m \times n$ $k$-layered cylindrical kirigami-based individual modules. As explained in the following results part, these multi-layered individual memory modules can perform with deformation sequences benefitted by gradient structural parameters in each segment. Superior than previous works (*35, 38, 39*) with just one-layer planar structural form, our designed mechanical memory device exhibit remarkable memory storage capability with the number of memory bit base exponentially increasing with segments layer quantity (Fig. 1B). We also demonstrate that our mechanical memory device can rather be feasibly fabricated by additive manufacturing technique (see the 3D printed prototype in Fig. 3C with $m = n = k = 3$).

To stably maintain the stored information in a robust way, we refer to the strong contraction force between the contrary "N" and "S" magnetic poles. Shown as Fig. 1D(i), we introduce permanent magnets into the non-cut part of the cylindrical kirigami modules. At the initial state (Fig. 1D(ii)), the force to buckle the kirigami segments $F_{Buckle}$ (Fig. 1E) is larger than the contraction force between two adjacent magnets. Therefore, the kirigami module can stand stably, i.e., stable state 1 (Fig. 1D(ii)). By appropriately selecting magnets with "N-S" poles contraction $F_{Contract}$ (Fig. 1E and inset) larger than the buckling force, we note the designed cylindrical kirigami module can stably hold its deformed state, i.e., stable state 2 shown as Fig. 1D(iii). Moreover, by purposely regulating the "N-S" poles contractions with large enough amplitude while still easily to be detached, our constructed mechanical memory device can rather robustly protect the stored information even under severe external interferences. We experimentally verify this bi-stable design concept by compressing the module with only one segment, and observe the negative force area in the compression-force curve (Fig. 1E) which is corresponding two energy minimums shown as Fig. 1F with $E_{min\_1} = 0$, $E_{min\_2} > E_{min\_1}$.

## RESULTS

### Tunable buckling behavior of one-layer cylindrical kirigami module

Shown as Fig. 2A(i-ii), we construct the cylindrical kirigami segment by introducing cut slits onto the cylindrical shell structure. We note that the buckling deformation feature (mainly buckling force) of this cylindrical kirigami module can be effectively tuned (Fig. 2B) just simply by varying the geometrical parameters including height H, width w and thickness t (Fig. 2A(ii)). Inspired by this observation and just via changing the heights of different structural segments (Fig. 2C(i)), we initially and successfully demonstrate a multi-layered cylindrical kirigami module that can deform sequentially shown as Fig. 2C(ii) and Fig. 2C(iii) wherein we use "0" and "1" to correspondingly represent the buckled and non-buckled segments. As demonstrated later, this deformation sequence will be the structural base to design multi-layered mechanical memory device with exceptional information storage capability. Note that to prevent the randomly directional buckling deformation of the uncut structural parts, we specially pre-assigned a curvature to induce them to buckle outward, see the 3D-printing manufactured prototype in Fig. 2A(iii).

By choosing distinct geometrical parameters, we systematically studied the mechanical performance of different cylindrical kirigami modules with experimental and numerical results shown in Fig. 2D-2H and fig. S1-S3. For the pre-curved uncut part exhibited as Fig. 2D, we can clearly see their uniform outward-buckling deformation features under compression (see movie S1-S2). Based on this result, we further explored the buckling deformations of cylindrical modules with divergent sets of geometrical parameters. Fig. 2E gives the displacement-force curves of kirigami modules separately with 10mm, 15mm and 20mm cut length while shell thickness maintains as T=1.2mm and the quantity of uncut part keeps as 12. Obviously, the design with larger height is with smaller buckling force which make it much easier to be buckled (i.e., with smaller buckling force) than those with smaller heights. Specifically, we note that under same and also the whole compression displacements/stage, the resist force of the structure with smaller height is always larger than that with larger height, i.e., $F_{smaller\_height} > F_{larger\_height}$. Moreover, we also observe that the initiating buckling position of the structure with larger height is always earlier than that with smaller structural height, see the comparison of experimental critical buckling strain in Fig. 2G with $(\varepsilon_b)_{H=10mm}$ (= 0.0853) < $(\varepsilon_b)_{H=15mm}$ (= 0.1041) < $(\varepsilon_b)_{H=15mm}$ (= 0.2313). We verify the experimental results by the consistent FEM simulation results (Fig. 2E, i to iii), and we believe the differences arise from the material defects and/or experimental set-up imperfections while within acceptable ranges.

Besides from structural heights, we also experimentally and numerically demonstrate that the buckling behavior of our proposed cylindrical kirigami structure can as well tuned through shell thickness. Fig. 2F gives the comparison results for three different cases with shell thickness T selected as 0.8mm, 1.2mm and 1.6mm while maintaining structural height as 20mm and uncut part number as 12. We note the design with larger shell thickness present not only with larger buckling force but also larger initiating buckling strain $\varepsilon_b$. Displayed as Fig. 2F, the buckling force for the case with shell thickness as 1.6mm is 41.16N which is about 1.5 times larger than the case with T=1.2mm and about 2 times larger than the case with T=0.8mm. However, compared with the strategy by tuning structural heights, the critical buckling strains under the thickness tuning method can also show obvious differences. For example, the value of $\varepsilon_b$ when T = 1.6mm is around 0. 12 which is 9.1% larger than the case when T=1.2mm with $\varepsilon_b$ =0.11, and 10% larger than the case when T=1.2mm with $\varepsilon_b$ =0.1.

It should note that the buckling deformation features do not show much obvious differences by tuning only the quantity of uncut parts (or number of introduced cut slits, see fig. S2), which is reasonable given the following equation (*44*)

$$F_{cr} \propto \frac{\pi^2 E T^2}{12(1-v^2)H^2}$$

with proof that the equivalent buckling force of the cylindrical kirigami module is deterministically scaled with both the shell thickness and length of the uncut parts, and wherein *E* and *v* are separately the material stiffness and Poisson's ratio. To comprehensively understand the buckling behavior of our proposed cylindrical kirigami structure, we numerically summarized the distributions of the buckling force under the coupling influence of the structural heights and shell thickness. Shown as Fig. 2H and fig. S3, we note the buckling force is always monotonically changing with the increments/decrements of heights and shell thickness, which, as demonstrated later, provides the basis for us to achieve and program deformation sequences in the multi-layered cylindrical kirigami modules.

**Sequential deformation of multi-layered cylindrical kirigami module**

Based on the results in last section, we demonstrate the sequential deformation behavior in multi-layered cylindrical kirigami module. Two strategies are verified, i.e., by assigning the multi-layered kirigami module only with gradient structural heights or only with variable shell thickness. We firstly explored the buckling behavior of the three-layered kirigami module by setting the heights of its three segments alternatively as 10mm, 15mm and 20mm (see Fig. 3A, left) while keeping the shell thickness as 1.2mm and uncut number as 12. Notably, we both experimentally and numerically observe the rather clearly three-staged deformation sequences in the 10-15-20 type of cylindrical kirigami (see the right in Fig. 3A and movie S3). And this sequential deformation feature is also reflected on its multi-staged displacement-force curve wherein the three buckling forces are relatively corresponding to the three different segments with different structural heights. Indicated by Fig. 3C, we know that the segment with largest structural height can buckle first and is with the smallest buckling force (experimentally, there is $(F_{Buckle})_{H=20mm}$ (~ 57.87N) < $(F_{Buckle})_{H=15mm}$ (~ 80.95N) < $(F_{Buckle})_{H=15mm}$ (~ 120.43N) ), which is consistent to the prediction by the above equation and the FEM based numerical results. Therefore, we can conclude that if the applied compression force F is lightly larger than $(F_{Buckle})_{H=20mm}$, the 10-15-20 type of module will only deform the 20mm segment whose deformed structural form is simply represented as (0,0,1) shown as structure ① in Fig. 3A. Meanwhile when $(F_{Buckle})_{H=20mm}$ < F < $(F_{Buckle})_{H=15mm}$, both the 20mm and 15mm segments will sequentially be buckled into the (0, 1, 1) type deformed structural form displayed as structure ② in Fig. 3A. Moreover, when $(F_{Buckle})_{H=10mm}$ < F, all the three segments will be sequentially buckled (i.e., the 20mm segment first, and followed by the 15mm segment and finally the 10mm segment, see movie S3), see ③ in Fig. 3A marked as (1,1,1).

Remarkably, we note the deformation sequence of the three-layered cylindrical kirigami module will not influenced by the different spatial combinations of its three divergent segments. To demonstrate, we compress the case with segments built as Fig. 3B, i.e., from top to bottom with order as (15, 10, 20). We find same deformation sequence occurs as the (10, 15, 20) design, i.e., the 20mm segment buckling first following with the 15mm segment and finally the 10mm segment. From the above equation, we can expect that these two designs should exhibit no difference in terms of their displacement-force curves. And we illustrate this based on the consistent numerical and experimental results displayed as Fig. 3C (the red and the blue lines). However, changing the spatial order of different segments can ultimately enable the cylindrical kirigami module engendering some novel deformed structural forms, such as the (1, 0, 1) type of sequentially buckled structure ② listed in Fig. 3B. Moreover, given the identical total segment length (i.e., sum($H_k$) = 45mm), we therefore can conclude that under same compression we can implement the deformation sequences to combinatorically encrypt a large number of divergent information, equivalently the memory storage bits, inside of the mechanical memory device. It is also the key why our proposed design concept can largely outperform previous one-layer works (*35, 38, 39*) which is constrained as one for each memory-storage structural element.

We also demonstrate the occurrence of deformation sequence for the three-layered cylindrical kirigami module programmed by using three different shell thickness. Shown as Fig. 3D-3E and movie S4, two sets of shell thickness are selected with spatial arrangement (from top to bottom) as (0.8, 1.2, 1.6) mm and (1.2, 0.8, 1.0) mm while maintaining height as 20mm and uncut number as 12. Four different deformed structural forms including (1,0,0), (0,1,0), (1,1,0) and (1,1,1), are observed. After being completely buckled to (1,1,1) state, both of the two cases exhibit three-staged displacement and force curves with little differences.

To further test the deformation sequence and also the mechanical behavior of our proposed multi-layered kirigami module, we printed a six-layered module shown as Fig. 3G (and more details about changing shell thickness and cut number in fig. S4-S6) and movie S5 whose segments are differentiated by assigning structural heights with order (from top to bottom) as (20,15,10,20,15,10)mm. Ideally with compression, three-staged deformation sequences similar as the three-layered case in Fig. 3A or 3B should occur, i.e. illustrated as the simulation and experimental results with only three different deformed structural forms: (1,0,0,1,0,0), (1,1,0,1,1,0) and (1,1,1,1,1,1), i.e., deformed structure states ②, ④ and ⑥ in Fig. 2G. However, impacted by the imperfections and/or the friction based resistance, we note that in experiments, some middle-staged structural forms are generated, for example deformed structures ①, ③ and ⑤ in Fig. 3G. Notably, these middle-staged structural state will totally not influence the deformation sequence of the three distinct structural segments, i.e. the two 15mm segment only initiate the buckling after the completion of two 20mm segments, and finally followed by the two 10mm segments (see movie S5). Interestingly, increasing the number of structural segments with same structural parameters can augment the damping area that are with almost same structural stiffness. For example, Fig. 3J compared the structural stiffness variations in terms of the compression displacement for the three-layered krigami module shown in Fig. 3A and 3B, and the six-layered case in Fig. 3G. Obviously, larger range of zero/small structural stiffness (almost two times of the two three-layered cases, see Fig. 3I) induced by the buckling of uncut parts are obtained in the six-layered kirigami module, which makes our proposed multi-layered kirigami structure ideal for energy absorption (Fig. 3J). However, in this work we focus only its potential applications in mechanical memory device with promising storage capability.

**Controllable mechanical memory storage capability**

Given the consistency between the numerical and the experimental results shown as Fig. 2 and Fig. 3, we further explored the controllable mechanical memory storage capability of single cylindrical kirigami module based on the FEA numerical methods. Displayed as Fig. 4A-4B and fig. S7-S8, the nine-layered module is selected for detailed analysis. By varying only the structural heights, we firstly study the designs with height spatial order (also from top to bottom, shown as Fig. 4A-4B and movie S6-S7) as $H_{1-9}$ = (10,15,20,10,15,20,10,15,20)mm (Fig. 4A(i)), (10,10,10,15,15,15,20,20,20)mm (Fig. 4A(ii)), (10,10,10,10,15,15,15,20,20)mm (Fig. 4B(i)), (10,10,15,15,15,20,20,20,20)mm (Fig. 4B(ii)), (10,15,15,20,15,20,10,15,20)mm (Fig. 4B(iii)), (20,10,10,15,20,10,15,15,20)mm (Fig. 4B(iv)). After fully compressing to trigger all segments buckling deformations, we find that for each case three deformed structural state occurring in sequence exist. For example, the case with $H_{1-9}$ = (10,15,20,10,15,20,10,15,20)mm can sequentially achieve (0,0,1,0,0,1,0,0,1), (0,1,1,0,1,1,0,1,1) and (1,1,1,1,1,1,1,1,1) modes while the case with $H_{1-9}$=(20,10,10,15,20,10,15,15,20)mm can sequentially deformed with (1,0,0,0,1,0,0,0,1), (1,0,0,1,1,0,1,1,1) and (1,1,1,1,1,1,1,1,1) modes. Illustrated as Fig. 4C and fig. S9, we compare the displacement-force curves of 10 different cases, and observe that all of them can display buckling-instability based mechanical responses and located into a square area. We note the multi-staged displacement-force curve can be found for the cases even that with identical segments (for example, $H_{1-9}$=(10,10,10,10,10,10,10,10,10)mm, (20,20,20,20,20,20,20,20,20)mm). We attribute this to the mutual influence of the buckling deformation among adjacent segments (see movie S7). However, according to how many types of segments are used, the cylindrical kirigami module can always uniquely achieve some sequentially deformed structural modes at certain forces $F_k$ (k is integer and equal to the number of segments types).

For example, for the studied multi-layer kirigami modules with 10mm-, 15mm- and 20mm-height types of segments, there are always three forces $F_1$, $F_2$ and $F_3$ (separately corresponding to the critical buckling force of 10mm, 15mm and 20mm segment with value around 60N, 81N and 128N when shell thickness set as 1.2mm and 12 uncut part) as the threshold to favor us determining which deformation sequence has been reached. Moreover, combining with the compression displacement shown as Fig. 4D, we can further ascertain the quantity of each type of segments after introducing permanent magnets with structural stability. For example, when the compression force F is in the range of [$F_1$, $F_2$) and the compression displacement reaches to 50mm, and meanwhile F is in the range of [$F_2$, $F_3$) with compression displacement as 95mm, we can derive that one (10, 15, 20) (no spatial order here) type of nine-layered structural module must be composed by five 10mm-height segments, three 15mm-height segment and one 20mm-height segment. Moreover, by setting each sequentially deformed structural state as one type of stored information pattern, we note the total number of stored information pattern of the (10, 15, 20) (no spatial order) segments based kirigami module will exponentially increase close to $2\times10^4$ different ones

(Fig. 4E). For generality, we find for any n-layered cylindrical kirigami module with $k$ ($k≥3$) type of different segments, there are always $k$ force thresholds to help us differentiate the sequential deformation modes. In fact, if equivalently treating the compression displacement as the stored information quantity, we note the multi-layered cylindrical kirigami module can as well increase the stored information density as explained explicitly in next section.

## Magnetically enabled mechanical memory device with robust structural stability and high information storage capability

Precedent meta-structure based mechanical memory storage devices either exhibit low information storage capability or are structurally too weak to resist external interferences. In last section, we have demonstrated the promising information storage capability of one single cylindrical kirigami module enabled by the sequential deformations of structural segments with gradient heights. In this section, we firstly illustrate how we robustly stabilize each buckled segment by referring to the strong contraction force of magnets "N-S" poles. Then, we prove the high information storage capability of our designed mechanical memory device by utilizing the deformation sequences and the combinatorics of the cylindrical kirigami modules. Shown as Fig. 5 and movie S8, we verify these through experiments and show how we can reprogram the information stored in our mechanical memory device by the erasing-and-rewriting process.

Fig. 5A gives the construction process of integrating permanent magnets into the two ends of each segment of a three-layered cylindrical kirigami module (from top to bottom as 22mm, 20mm and 18mm segments). The magnets are fastened into the module by using their slightly larger diameter dimensions than the module and also the glue pasting to generate strong enough friction force. Note that the magnets are placed with identical posture to guarantee N-S poles contractions. The key to choose appropriate magnets relies on the truth that the N-S poles contraction force is not larger than the buckling force of each segments to maintain the stability (i.e. stable state 1) of the module at the initial free loading condition while strong enough to resist the plateau buckling force of the cylindrical kirigami module after contracting together. By choosing the suitable magnet, we experimentally test the mechanical performance of three kirigami modules with heights separately as (18, 20, 22)mm, (18, 22, 20)mm and (20, 18, 22)mm while shell thickness all as 1.2mm and the number of uncut parts as 8. The test results are shown in Fig. 5B. Based on the displacement-control compression method, we clearly see the negative force area which stands for the N-S poles contraction. Moreover, we find all the tested three cases display similar three-staged displacement-force curves, and the contraction forces are always larger than the plateau buckling force of each segment. For example, for the (18, 22, 20) case, the N-S poles contraction forces after the buckling of each segment are separately around −10.004N, −7.703N and −8.152N which are larger than their corresponding plateau buckling forces relatively as 6.621N, 5.988N and 3.647N (Fig. 5C). According to the contraction force and the structural dimensions, we evaluate the structural robustness to resist external interruptions by calculating the structural stiffness of the kirigami module at the second stable state. We find its structural stiffness can reach to 10KPa which is 100 times larger than previously reported works (*38*).

Fig. 5D gives one three-layered prototype integrated with four permanent magnets and the process that it uses deformation sequence to store different information patterns. Fig. 5D(i) lists all eight potential information patterns that the three-segment cylindrical kirigami module can stored. Truthfully, each cylindrical kirigami module can mostly store three different information patterns, see the information pattern transition from (0,0,0) to (0,1,0), then to (0,1,1) and finally settle at (1,1,1) highlighted in Fig. 5D(i) and its corresponding experimental demonstrations in Fig. 5D(ii-iii). We note this multi-staged information transition has superiorly outperform the tradition memory bits which can only transform between two states (generally representing as "0" and "1"). During experiments, even large force needed, we still can erase the already patterned information (see the represented "write" process from (0,0,0), to (1,1,1) bypassing states (0,1,0) and (0,1,1) in Fig. 5D(ii)) by stretching the module inversely (see the pattern transition process from (1,1,1), to (0,0,0) bypassing states (0,1,1) and (01,0) in Fig. 5D(iii)). We note that this "write-erase" and then "rewrite" features also provides the basis for our following demonstrated mechanical memory device to easily transform among any different information patterns.

Fig. 5E(i) and (iii) (see more details in fig. S14) display the overview of our proposed mechanical memory storage device. It is constructed by placing identical cylindrical kirigami modules into an m×n array (3×3 in Fig. 5E with three-layered kirigami module demonstrated in Fig. 5D). Note that all these kirigami modules can independently achieve their corresponding sequential deformations. Then, we can use the combinatorics of their sequentially deformed patterns to store a very large number of information. However, beyond traditional mechanical memory designs to use in-plane pixel-like structural element as memory bit, the sequential break this limit and extend the memory bit from planar form into spatial combinatorics. Therefore, the storage capability have increase exponentially as we initially present in Fig. 1B. To demonstrate, we list 25 (see fig. S15 and movie S9) representative achievable information patterns (totally as $3^9$) of one 3×3 design with initial information pattern as [(0,0,0), (0,0,0), (0,0,0); (0,0,0), (0,0,0), (0,0,0); (0,0,0), (0,0,0), (0,0,0)] (Fig. 5E(ii)). Following certain combinatorically "write" paths, this 3×3 mechanical memory device can purposely be deformed into certain patterns to stand for unique information. For example, we give one write path in Fig. 5F(i), i.e. #1-#2-#3-#4-#12-#16-#25, and several other information patterns in Fig. 5F(ii).

To test the robustness of the deformed structure after storing with certain information, we artificially introduce some external interference including mechanical crushing and shaking onto the structure. We find the structure can stand rather stably without any "damage" or "loss" for the stored information bits (see movie S10).

**DISCUSSION**

In summary, this work propose a new strategy to design stable mechanical memory devices with both ultra-high information storage capability and promising structural robustness. The cylindrical kirigami shell module with multiple segments of gradient structural parameters exhibits controllable deformation sequences which makes it equivalently generate memory bits beyond the traditional "0" and "1" states. We create the mechanical memory device by combining certain number of multi-segment kirigami modules together. Both the deformation sequences, corresponding to memory bits, and the deformation independence of the kirigami modules enable our proposed mechanical memory device can store information exponentially increasing with the number of kirigami module and its composed segments. The introduced "N-S" magnetic poles provides desired contraction forces to robustly stabilize the stored information. The actuation method is not included in this work given its simplicity, which can be achieved by any artificial methods for non-remotely control and magnetic way for remote and untouched control. Compared with precedent works, our work has several merits: (1) ultra-high information storage capability; (2) high structural stability robust to resist severe external interference; (3) simple structural form and feasible to be fabricated just through 3D printing techniques. Introducing the gradient structural design concept and referring to magnetic poles for achieving structural stability on intrinsically buckling-instability based structure pave a new way to design mechanical meta-structures/metamaterials with unexpected properties and functionalities.

**MATERIALS AND METHODS**

**Fabrication details**

All prototypes are manufactured with TPU materials by FDM 3D printer (F900 Industrial 3D Printer, Stratasys). The ratio between cut slits and uncut part is 1:5. We fixed the bridge part among cut segments as 10mm and internal radius of the prototype as 10mm. One solid cylindrical column and one cylindrical shell (both are with length as 101.2mm for three-segmented kirigami module, and 191.2mm for six-segmented module while their shell thickness as 1.2mm, see fig. S10b) are also printed with PLA material by the same FDM 3D printer (F900 Industrial 3D Printer, Stratasys) for compression experimental uses. Prototypes with three segments (one with 10mm, 15mm and 20mm segments, and the second one with

18mm, 20mm and 22mm segments, see Fig. 2) and also six segments (see details in Fig. 3) are fabricated correspondingly.

Additionally, the mechanical memory devices are made by 3 by 3 three-segmented cylindrical kirigami modules. The solid column part is inserted into each module to maintain the vertical compression while at the top end the cylindrical shell is pasted by ultra-adhesive glue (Gorilla Super Glue Gel). Round Disc Donut Magnets (X-bet MAGNET) are used with sizes internal radius slightly larger than the inner radius of the printed sample. To further fasten the magnet and prevent any movements during compression, glue is also used when placing the magnet into the kirigami module.

**Experimental methods**

During compression experiments, the printed solid column is used to support the gradient cylindrical krigami and make it buckle always vertically under compression and the printed cylindrical shell is used for compressing the cylindrical kirigami sample (fig. S11). The GOTECH TCS-2000-GDL machine is implemented to vertically compress the sample and obtain the displacement-force curves. The compression rate is set as 50mm/min and the compressing displacement is equal to the total length of the sample segments (for example, the (10, 15, 20)mm type of sample with compression displacement as 10mm+15mm+20mm=45mm).

**FEM simulation details**

The commercial simulation software Dassault Systems Simulia ABAQUS is used and under the Explicit mode. The cylindrical kirigami module is simulated as 3D shell model while the support cylindrical shell is simulated as 3D solid model. The material parameters of the TPU based kirigami modules are determined through uniaxial compression experiments (fig. S10a) while the PLA based support part is experimentally tested as 4.107GPa and Poisson's ratio is assumed as 0.35. Explicit solver is used for buckling analysis. The S4R element is used for kirigami module while the C3D8R element is used for the support cylindrical shell. The bottom boundary of the cylindrical support shell is totally constrained while the top end of the kirigami module is compressed with displacement along with its axial direction (fig. S12). To imitate the contact between the internal solid column and the kirigami module, the contact condition is set with penalty parameter as 0.1. Moreover, elements sizes (0.4mm, 0.5mm, 1mm, 1.5mm and 2mm) are selected for simulation results comparison (fig. S13).

**Supplementary Materials**

**This PDF file includes:**

Notes S1 to S3

Figs. S1 to S15

Legends for movies: S1 to S10

**Other Supplementary Materials for this manuscript includes the following:**

Movie S1 to S10

**Acknowledgement: Funding**: This work was supported by National Natural Science Foundation of China (No. 11972144), Shanxi Province Specialized Research and Development Breakthrough in Key Core and Generic Technologies (Key Research and Development Program) (No. 2020XXX017), and Fundamental Research Program of Shanxi Province (No. 202203021211134), which are gratefully acknowledged. **Author Contributions**: Y. L., L. X. proposed and designed the research project. B. W., L. X, conducted the experiments and simulations. Y. L., L. X. analyzed the data and composed the figures. Y. L. wrote the paper. All of the authors discussed the results and edited the manuscript. **Competing interest**: There is no competing interests for all the authors. **Data and materials availability**: All data needed to evaluate the conclusions in the paper are present in the paper and/or in the Supplementary Materials.

# FIGURES

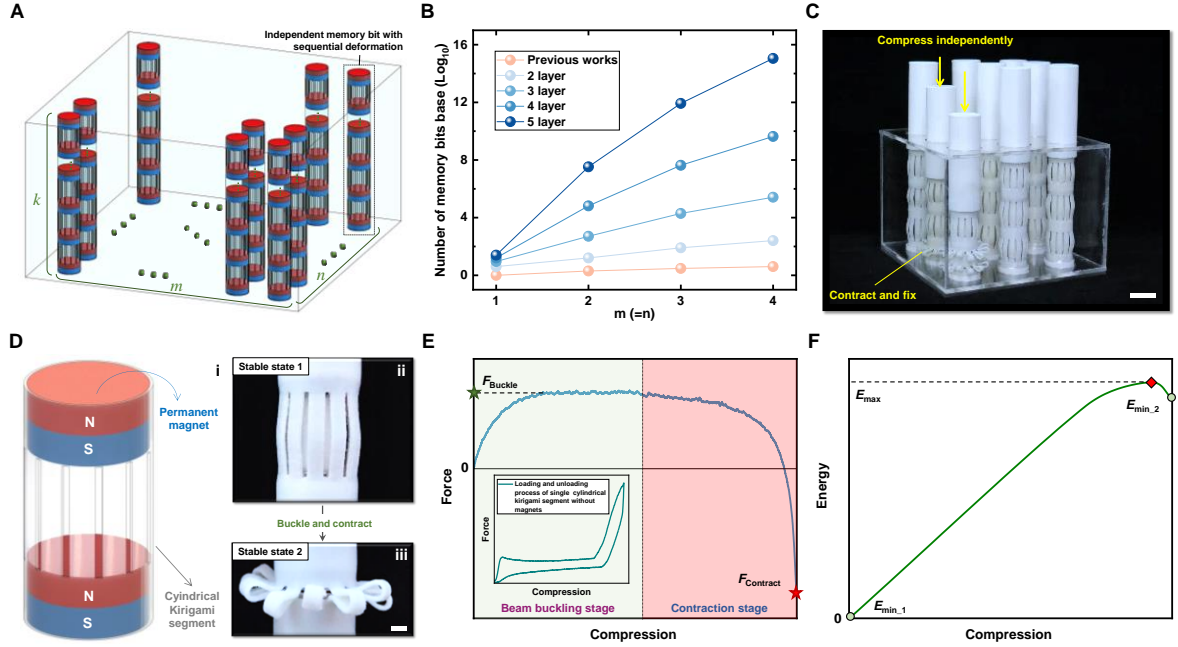

**Fig. 1. The overview of the proposed mechanical memory devices**. (A) Schematics of our constructed mechanical memory devices composed by $m \times n$ $k$-layered cylindrical kirigami modules integrated with permanent magnets. (B) The high information storage capability of our proposed mechanical memory devices demonstrated through the relation between the number of information memory bits and the unit number. (C) The prototype of our built mechanical memory device made by 3D printed cylindrical kirigami units. Scale bar: 10mm (D) Illustration of the achievement of structural stability (stable state 1 in (ii) and stable state 2 in (ii)) in one-layered cylindrical kirigami module by (i) introducing magnetic poles. Scale bar: 2mm (E) Structural bi-stability demonstration of one-layered kirigami module with magnetic poles through uniaxial compression test. (F) The energy map of the compressed one-layered kirigami module in (E) wherein clearly two energy minimums $E_{\min\_1}$ and $E_{\min\text{-}2}$ can be observed.

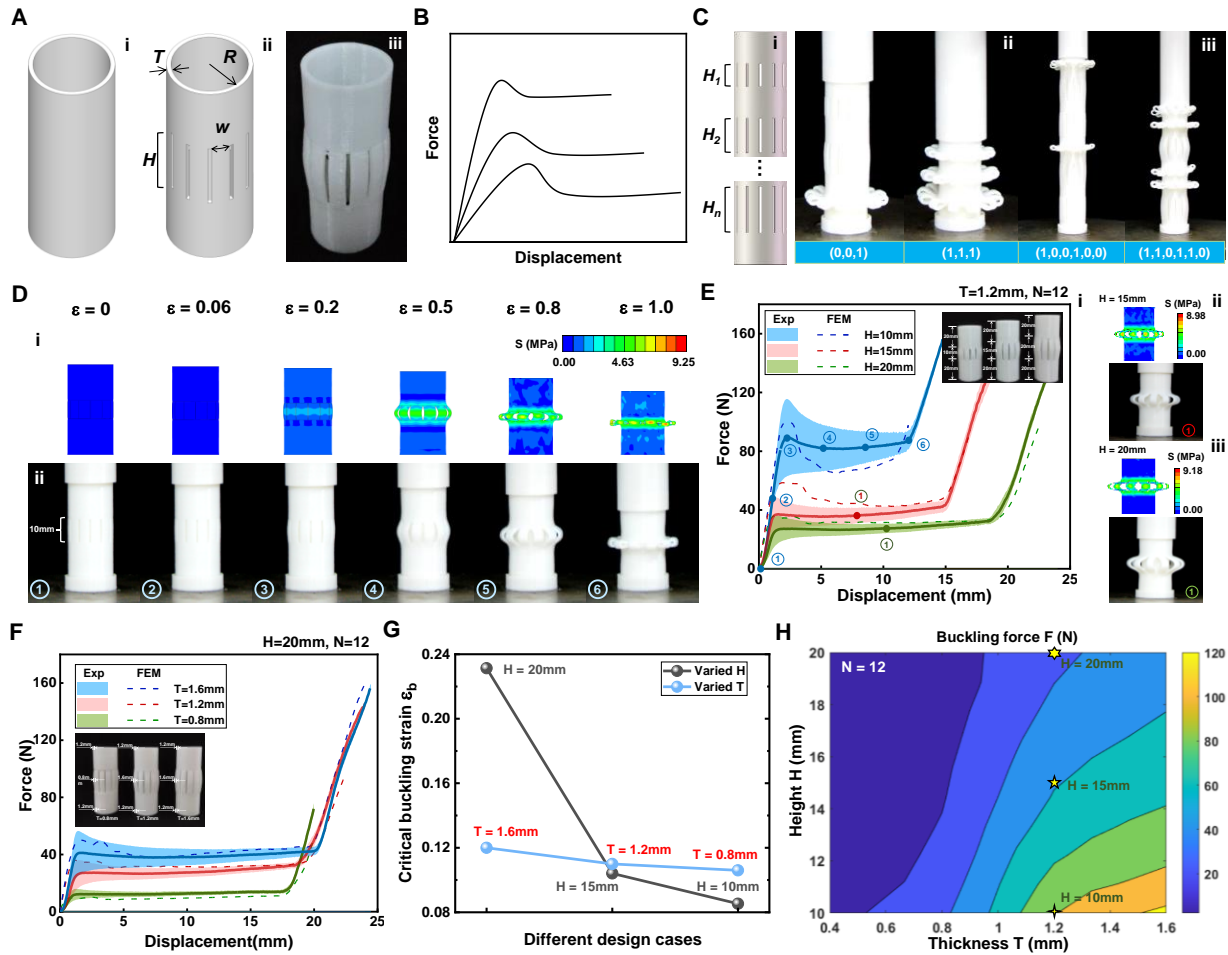

**Fig. 2. Illustration of tunable buckling instability of our proposed one-layer cylindrical kiriagmi module**. (**A**) Illustration of the proposed one-layered cylindrical kirigami module (i to ii) printed with pre-assigned curvature on uncut parts (iii). (**B**) Illustration of tunable buckling behavior (force and displacement curve) of curved beam by programming structural parameters. (**C**) Overview of buckled cylindrical kirigami module with deformation sequences: schematic illustration by designing with linearly varied structural height cut slits (i), and the sequentially deformed prototypes with three segments (ii) and six segments (iii). (**D**) Experimental and numerical buckling behavior of the proposed cylindrical kirigami module with one segment. (**E-F**) Illustration of tunable buckling behavior of the cylindrical kirigami module through gradient cut length (E) and the shell thickness (F). (**G**) Comparisons of buckling behavior by tuning cut length (the height of uncut part) and the shell thickness. (**H**) The comprehensive exploration of the buckling behavior through critical buckling force continuously changing with shell thickness T and cut height H.

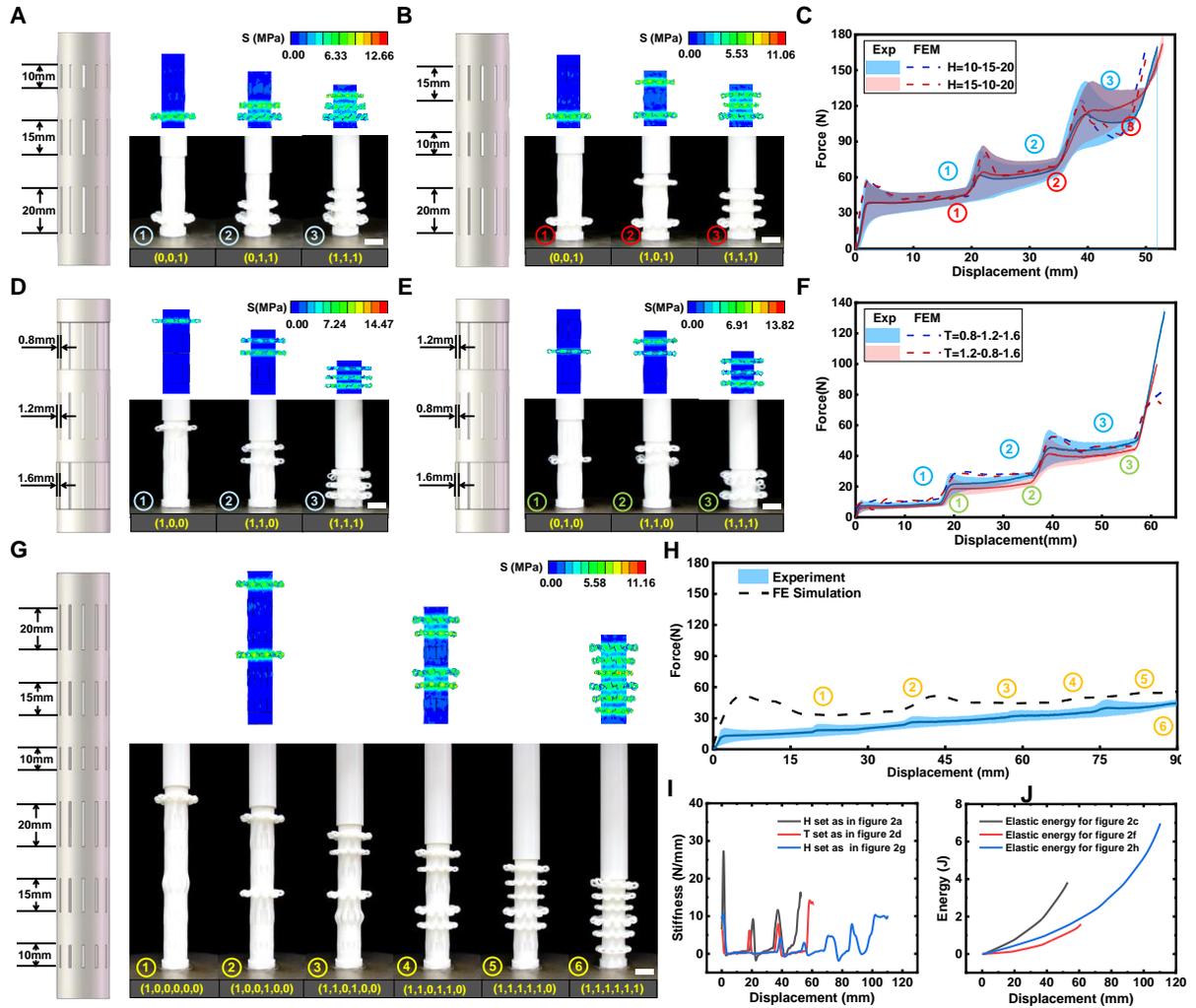

**Fig. 3. Deformation sequences of multi-layered cylindrical kirigami module programmed by gradient structural heights (length of cut slits) or shell thickness.** (**A-B**) Experimental and numerical demonstration of the deformation sequences of three-layered cylindrical kirigami tuned by cut length: (10, 15, 20)mm case in (A) and (15,10,20)mm case in (B). (**C**) Comparison of displacement and force compression curves of the three-layered kirigami module in (A) and (B). (**D-E**) Experimental and numerical demonstration of the deformation sequences of three-layered cylindrical kirigami tuned by shell thickness: (0.8, 1.2, 1.6)mm case in (D) and (1.2,0.8,1.6)mm case in (E). (**F**) Comparison of displacement and force compression curves of the three-layered kirigami module in (D) and (E). (**G**) Experimental and numerical demonstration of the deformation sequences of six-layered cylindrical kirigami tuned by cut length: (20, 15, 10, 20, 15, 10)mm. (**H**) Displacement and force compression curves of the six-layered kirigami module in (H). (**I**) Comparison of structural stiffness of the three- and six-layered cylindrical kirigami module in (A, B, D, E, G). (**J**) Comparison of stored energy in the kirigami module in (A, D, G).

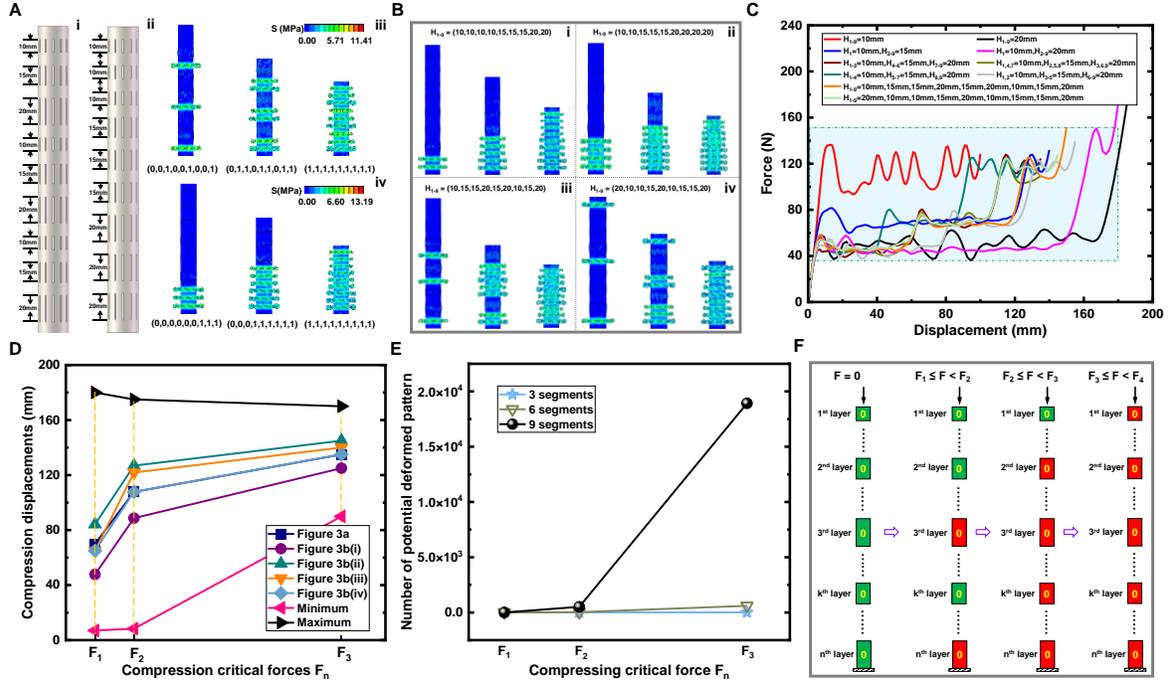

**Fig. 4. Controllable mechanical memory storage capability of single cylindrical kirigami module.** (**A**) FEM results of the deformation sequences of two nine-layered cylindrical kirigami modules: (10,15,20,10,15,20,10,15,20)mm in (i) and (10,10,10,15,15,15,20,20,20)mm in (ii). (**B**) FEM results of the deformation sequences of some other representative nine-layered cylindrical kirigami modules: (10,10,10,10,15,15,15,20,20)mm in (i), (10,10,15,15,15,20,20,20,20) in (ii), (10,15,15,20,15,20,10,15,20) in (iii) and (20,10,10,15,20,10,15,15,20) in (iv). (**C**) Numerical exploration of mechanical behaviors of randomly designed nine-layered cylindrical kirigami modules through the displacement and force curve. (**D**) Differentiating of nine-layered cylindrical kirigami module composed by (10,15,20) type segments through the relations between the compression critical force $F_n$ (n=1,2,3) (force threshold) and the related compression displacement. (**E**) Relations between the number of the potential deformed pattern of three-/six-/nine-layered cylindrical kirigami modules composed by (10,15,20) type segments with the compression buckling force $F_n$. (**F**) Schematic illustration of the controllable deformed pattern of n-layered cylindrical kirigami modules composed by k different segments corresponding with the related critical buckling force $F_k$ (k=1,2,3,…).

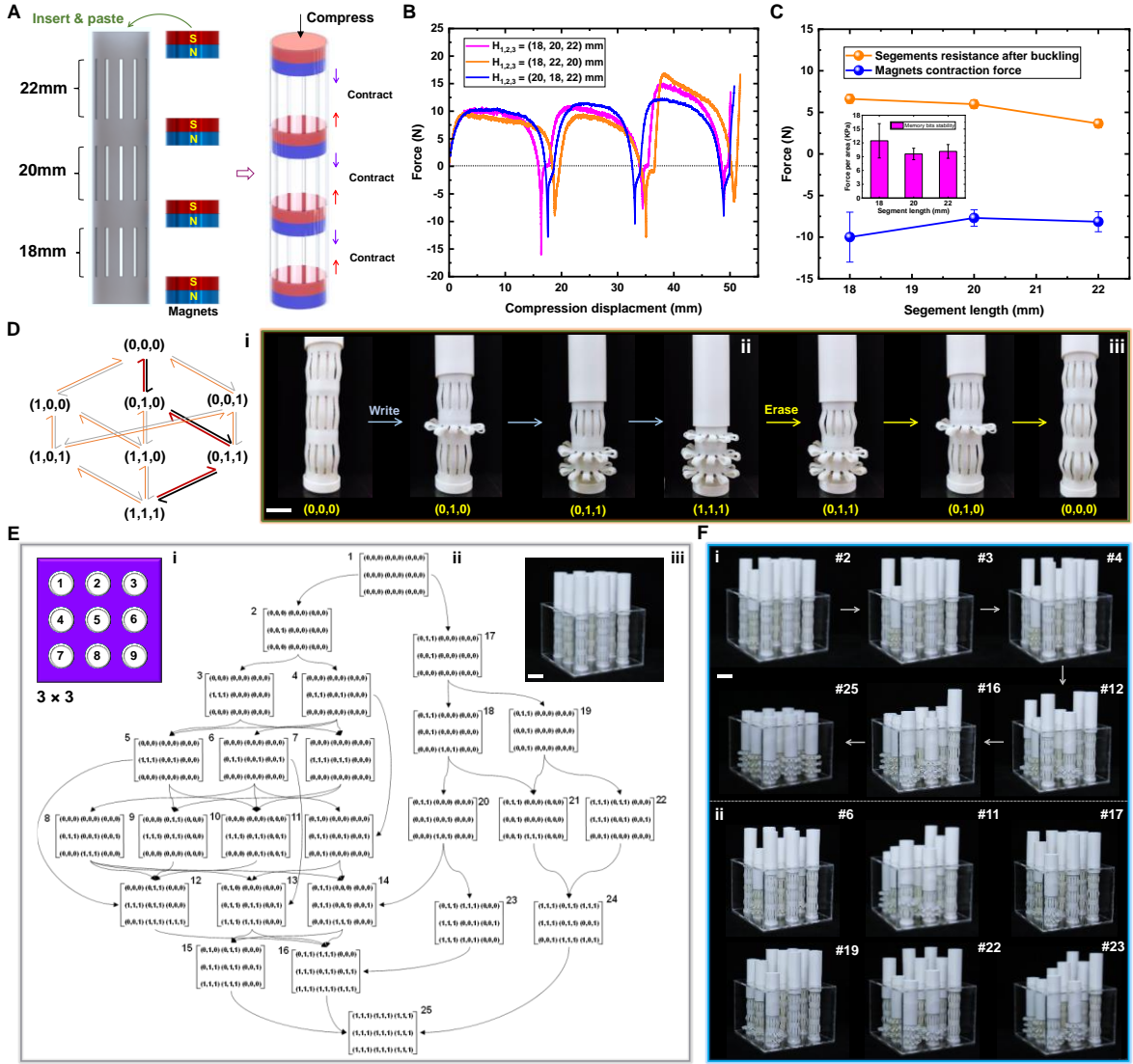

**Fig. 5. The gradient cylindrical kirigami module composed mechanical memory device with high information storage capability.** (**A**) Schematic illustration of the integration of magnetic poles (disk shaped permanent magnets with poles direction along axial direction) into a three-layered cylindrical kirigami module with (22,20,18)mm type of segments. (**B**) Experimental compression test of three cylindrical kirigami modules ((18,20,22), (18,22,20) and (20,18,22)) combining with magnetic poles. (**C**) Comparison between the maximum poles contraction force and the buckling force of different segments in (A). (**D**) Demonstration of stable deformation sequences of three-layered kirigami module composed by 18mm, 20mm and 22mm segments. (**E**) Illustration of the proposed mechanical memory device composed by 3 by 3 identical modules (i) and (iii), and their representative reprogrammable deformation sequences (or equivalently the type of information patterns). (F) Experimental demonstration of the reprogrammable information storage capability of the 3 by 3 mechanical memory device in (E): the deformation sequence along branch #2-#3-#4#-#12-#16-#25 (i) and some randomly selected deformed patterns.